\documentclass[twocolumn]{emulateapj}

\usepackage{amsmath}
\usepackage{physics}
\usepackage{graphicx}
\usepackage{color}
\usepackage[colorlinks]{hyperref}
\hypersetup{
    colorlinks,	
    citecolor=blue,
}

\newcommand{\be}{\begin{eqnarray}}
\newcommand{\ee}{\end{eqnarray}}

\shortauthors{Mirzaev et al.}

\begin{document}

\title{X-ray spectra of black hole X-ray binaries with returning radiation}

\author{Temurbek~Mirzaev\altaffilmark{1}, Cosimo~Bambi\altaffilmark{1,2,\dag}, Askar~B.~Abdikamalov\altaffilmark{2,3,1}, Jiachen~Jiang\altaffilmark{4,5}, Honghui~Liu\altaffilmark{1,6}, Shafqat~Riaz\altaffilmark{6}, and Swarnim~Shashank\altaffilmark{1}}

\altaffiltext{1}{Center for Astronomy and Astrophysics, Center for Field Theory and Particle Physics, and Department of Physics,
Fudan University, Shanghai 200438, China. \email[\dag E-mail: ]{bambi@fudan.edu.cn}} 
\altaffiltext{2}{School of Natural Sciences and Humanities, New Uzbekistan University, Tashkent 100007, Uzbekistan}
\altaffiltext{3}{Ulugh Beg Astronomical Institute, Tashkent 100052, Uzbekistan}
\altaffiltext{4}{Institute of Astronomy, University of Cambridge, Madingley Road, Cambridge CB3 0HA, UK}
\altaffiltext{5}{Department of Physics, University of Warwick, Gibbet Hill Road, Coventry CV4 7AL, UK}
\altaffiltext{6}{Institut f\"ur Astronomie und Astrophysik, Eberhard-Karls Universit\"at T\"ubingen, D-72076 T\"ubingen, Germany}

\begin{abstract}
In the disk-corona model, the X-ray spectrum of a stellar-mass black hole in an X-ray binary is characterized by three components: a thermal component from a thin and cold accretion disk, a Comptonized component from a hot corona, and a reflection component produced by illumination of the cold disk by the hot corona. In this paper, we assume a lamppost corona and we improve previous calculations of the X-ray spectrum of black hole X-ray binaries. The reflection spectrum is produced by the direct radiation from the corona as well as by the returning radiation of the thermal and reflection components and is calculated considering the actual spectrum illuminating the disk. If we turn the corona off, the reflection spectrum is completely generated by the returning radiation of the thermal component, as it may happen for some sources in soft spectral states. After choosing the radial density profile of the accretion disk, the ionization parameter is calculated self-consistently at any radial coordinate of the disk from the illuminating X-ray flux and the local electron density. We show the predictions of our model in different regimes and we discuss its current limitations as well as the next steps to improve it.   
\end{abstract}


\section{Introduction}

X-ray reflection spectroscopy is the analysis of relativistically blurred reflection features in the X-ray spectra of stellar-mass black holes in X-ray binary systems and supermassive black holes in active galactic nuclei~\citep{2014SSRv..183..277R,2021SSRv..217...65B}. It is a powerful tool to study the accretion process in the strong gravity region of these systems, measure black hole spins \citep[see, e.g.,][]{2006ApJ...652.1028B,2023ApJ...946...19D,2023arXiv231116225D}, and test Einstein's theory of general relativity in the strong field regime \citep[see, e.g.,][]{2017ApJ...842...76B,2019ApJ...875...56T,2021ApJ...913...79T}. The past 10-15~years have seen a remarkable progress in our capability of analyzing these relativistically blurred reflection features, thanks to new X-ray observatories and more advanced theoretical models~\citep{2004ApJS..153..205D,2014ApJ...782...76G,2019MNRAS.485.2942N,2019MNRAS.488..324I,2019ApJ...878...91A}. However, current theoretical models still rely on a number of simplifications, so caution is necessary when we analyze high-quality data and we get very precise measurements. Moreover, the next generation of X-ray missions promises to provide unprecedented high-quality data that will necessarily require more advanced synthetic spectra than those available today.

\begin{figure}[t]
\begin{center}
\includegraphics[width=0.95\linewidth]{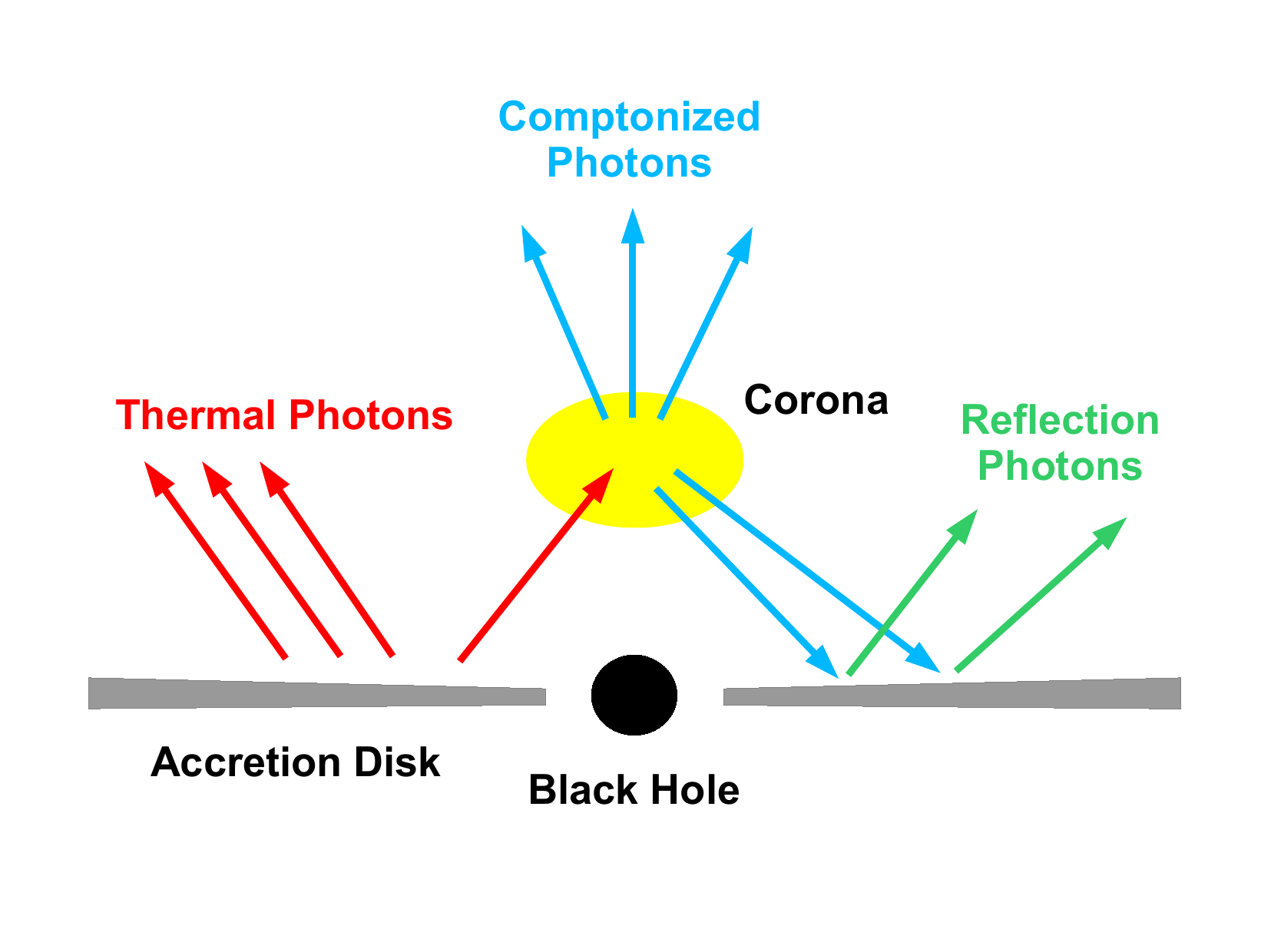}
\end{center}
\vspace{-1.0cm}
\caption{Disk-corona model. Figure from \citet{Bambi:2021chr} under the terms of the Creative Commons Attribution 4.0 International License. \label{f-corona}}
\end{figure}

In the disk-corona model, we have a black hole, a cold accretion disk, and a hot corona, as shown in Fig.~\ref{f-corona}. The black hole can be either a stellar-mass black hole in an X-ray binary or a supermassive black hole in an active galactic nucleus. The disk is geometrically thin and optically thick, and it is ``cold'' because it can efficiently emit radiation. The thermal spectrum of the disk is peaked in the soft X-ray band (0.1-10~keV) for stellar-mass black holes in X-ray binaries and in the UV band (1-100~eV) for supermassive black holes in active galactic nuclei. The corona is some ``hot'' plasma ($\sim 100$~keV) near the black hole and the central part of the accretion disk. For example, the base of a jet, the atmosphere above the accretion disk, or the material in the plunging region between the inner edge of the disk and the black hole may act as a corona. Two or more coronae may also coexist at the same time. Since the disk is cold and the corona is hot, thermal photons from the disk (red arrows in Fig.~\ref{f-corona}) can inverse Compton scatter off free electrons in the corona. The X-ray spectrum of the Comptonized photons (blue arrows in Fig.~\ref{f-corona}) can be normally approximated by a power law with a high-energy cutoff. A fraction of the Comptonized photons can illuminate the disk: Compton scattering and absorption followed by fluorescent emission generate the reflection spectrum (green arrows in Fig.~\ref{f-corona}).

In the rest-frame of the material of the disk, the reflection spectrum is characterized by narrow fluorescent emission lines in the soft X-ray band and a Compton hump with a peak around 20-40~keV~\citep{2005MNRAS.358..211R,2010ApJ...718..695G}. The most prominent emission feature is usually the iron K$\alpha$ complex, which is at 6.4~keV in the case of neutral or weakly ionized iron atoms and shifts up to 6.97~keV in the case of H-like iron ions. The reflection spectrum of the whole disk detected far from the source is blurred because photons emitted from different points of the disk arrive at the detector with a different redshift, which is the combination of a gravitational redshift to climb the gravitational well of the black hole and a Doppler boosting due to the motion of the material in the disk \citep{2017bhlt.book.....B}.

There is a relatively rich literature on the study of the simplifications in current reflection models and on their impact on current measurements of accreting black holes, as well as attempts to remove such simplifications. Simplifications in current reflection models mainly appear in the structure of the accretion disk, in the description of the corona, and in the calculations of the reflection spectrum in the rest-frame of the material in the disk. Even some relativistic effects are not properly taken into account. For a non-complete list of these studies, see, for instance, \citet{2008MNRAS.386..759N,2011MNRAS.414.1269W,2012MNRAS.424.1284W,2016ApJ...821L...1N,2017MNRAS.472.1932G,2018MNRAS.477.4269N,2020PhRvD.101l3014C,2020ApJ...895...61R,2021ApJ...910...49R,2020ApJ...899...80A,2021ApJ...923..175A,2022ApJ...938...53S,2022MNRAS.517.5721M,2024ApJ...965...66M}.

The {\it returning radiation} (or self-irradiation) is the radiation emitted by the disk and returning to the disk because of the strong light bending near a black hole. \citet{1997MNRAS.288L..11D} were the first to study the returning radiation in the context of disk reflection, concluding that the effect is not important when the corona corotates with the disk. Such a conclusion was confirmed in \citet{2008MNRAS.386..759N}, where it was also showed that the effect is instead important if the corona is static and close to the black hole. \citet{2016ApJ...821L...1N} and \citet{2018MNRAS.477.4269N} studied the returning radiation in the context of lamppost coronae and showed the effect can be important and non-negligible in certain situations.  The impact of the returning radiation on reflection spectra has been recently investigated in a few studies with different approximations~\citep{2020MNRAS.498.3302W,2022MNRAS.514.3965D,2021ApJ...910...49R,2023arXiv230312581R}. \citet{2022MNRAS.514.3965D} implemented the effect of the returning radiation in the {\tt relxill} model by introducing a modified emissivity profile but still calculating the reflection spectrum in the rest-frame of the material in the disk assuming that the illuminating X-ray flux has a spectrum described by the power law with a high-energy cutoff.

In~\citet{2024ApJ...965...66M}, we showed that the returning radiation can have a strong impact on the reflection spectrum for some systems and its effect cannot be reabsorbed by a modified emissivity profile with respect to the one produced by the direct radiation from the corona, because the spectrum from the corona is approximately a power law with a high-energy cutoff while the spectrum of the returning radiation is a reflection spectrum. The effect is particularly strong when the inner edge of the disk is very close to the black hole, the corona illuminates mainly the inner part of the accretion disk, and in the case of a low or moderate value of the ionization parameter. For such systems, it is crucial to calculate the reflection spectrum at every radial coordinate of the disk by using the actual spectrum of the incident radiation. If we do not do so, the model cannot fit the data well and some parameters can be significantly overestimated or underestimated \citep{2024ApJ...965...66M}.

In this manuscript, we extend the work in \citet{2024ApJ...965...66M} and we present the model {\tt ziji}, which is available on GitHub\footnote{\url{https://github.com/ABHModels/ziji/}.} and Zenodo~\citep{ziji-zenodo}. {\tt ziji} assumes a lamppost corona and calculates the thermal spectrum of the cold disk, the Comptonized spectrum from the corona, and the reflection spectrum from the disk. The reflection spectrum is produced by the direct radiation from the corona as well as by the returning radiation of the thermal and reflection components and is calculated considering the actual spectrum illuminating the disk. We note that current public models assume that the radiation illuminating the disk is either a power law/Comptonized spectrum \citep[like {\tt xillver},][]{2013ApJ...768..146G} or a single-temperature blackbody spectrum \citep[like {\tt xillverNS},][]{2022ApJ...926...13G}, but there is no model that considers the possibility that the radiation illuminating the disk is some combination of these two components. Moreover, the returning radiation of the thermal component is not a single-temperature blackbody spectrum but a multi-temperature blackbody spectrum whose exact shape is different at different radii because the returning radiation illuminating a certain point is the combination of the returning radiation from different parts of the accretion disk, with different temperature and different redshift. Our model also includes higher-order reflections, because even the reflection spectrum can return to the disk and produce reflection, and this further changes the actual spectrum of the incident radiation and, in turn, of the reflection component. In \citet{2024ApJ...965...66M}, we ignored the thermal spectrum of the disk and therefore there was no returning radiation of the thermal component in the incident spectrum generating the reflection spectrum. We note that the new calculations presented here can be important only for stellar-mass black holes in X-ray binaries, as their thermal spectrum can be peaked in the soft X-ray band and it can thus have an impact on the reflection spectrum in the X-ray band. In the case of supermassive black holes in active galactic nuclei, the thermal spectrum is peaked in the UV band and its returning radiation cannot appreciably affect the reflection spectrum in the X-ray band. We also note that there is some observational evidence that the reflection spectrum can be produced mainly by the returning radiation of the thermal component when a source is in the soft spectral state, where the total spectrum is dominated by the thermal component and the coronal spectrum is weak or absent \citep{2021ApJ...909..146C}.

The manuscript is organized as follows. In Section~\ref{s-mod}, we describe the assumptions of our new model. In Section~\ref{s-res}, we present the predictions of {\tt ziji} in different regimes (when the spectrum of the source is dominated by the Comptonized spectrum of the corona, when it is dominated by the thermal spectrum of the accretion disk, and when these two components are roughly equivalent). In Section~\ref{s-dis}, we discuss our results, current limitations of {\tt ziji}, and future steps to improve our model. Throughout the manuscript, we assume that the spacetime metric has signature $(-+++)$ and employ natural units in which $c = G_{\rm N} = \hbar = k_{\rm B} = 1$.


\section{Model}\label{s-mod}

The default spacetime metric in our model is the Kerr solution. However, {\tt ziji} does not use any specific equation or property of the Kerr metric and we can consider any stationary, axisymmetric, and asymptotically flat spacetime simply by changing the metric coefficients as long as the metric is written in spherical-like coordinates $(t,r,\theta,\phi)$, the only non-vanishing off-diagonal metric coefficient is $g_{t\phi}$ (circular spacetime), and there are no pathological properties (like naked singularities, which would require some special treatment to avoid problems in the numerical calculations).

The accretion disk is described by the Novikov-Thorne model \citep{1973blho.conf..343N,1974ApJ...191..499P}. The disk is infinitesimally thin, perpendicular to the black hole spin axis, and the inner edge of the disk is at the innermost stable circular orbit (ISCO) or at a larger radius. The material in the disk moves on nearly-geodesic equatorial circular orbits. From the conservation of mass, energy, and angular momentum, the model predicts the time-averaged radial structure of the disk. The time-averaged energy flux emitted from the surface of the disk at the radial coordinate $r$ is \citep{1974ApJ...191..499P,2017bhlt.book.....B}
\be
\mathcal{F}(r) = \frac{\dot{M}}{4 \pi M^2} F(r) \, ,
\ee
where $M$ is the black hole mass, $\dot{M} = dM/dt$ is the mass accretion rate and is independent of $r$ (no outflows/winds), and $F(r)$ is the following dimensionless function
\be\label{eq-NT}
F(r) = - \frac{\partial_r \Omega}{\left( E - \Omega L_z \right)^2} \frac{M^2}{\sqrt{-G}} \int_{r_{\rm in}}^{r} \left( E - \Omega L_z \right) \left( \partial_\rho L_z \right) d\rho \, . \nonumber\\
\ee
$G$ is the determinant of the near equatorial plane metric, $r_{\rm in}$ is the radius of the inner edge of the accretion disk ($r_{\rm in} \ge r_{\rm ISCO}$), and $\Omega$, $E$, and $L_z$ are, respectively, the angular velocity, the conserved specific energy, and the conserved specific angular momentum of the material in the disk \citep[see, e.g.,][]{2017bhlt.book.....B}
\be
\Omega &=& \frac{-\partial_r g_{t\phi} \pm 
\sqrt{ \left(\partial_r g_{t\phi}\right)^2 - \left(\partial_r g_{tt}\right)\left(\partial_r g_{\phi\phi}\right) } }{\partial_r g_{\phi\phi}} \, , \\
E &=& - \frac{g_{tt} - \Omega g_{t\phi}}{\sqrt{-g_{tt} - 2 \Omega g_{t\phi} -\Omega^2 g_{\phi\phi}}} \, \label{eq-energy} \\
L_z &=& \frac{g_{t\phi} - \Omega g_{\phi\phi}}{\sqrt{-g_{tt} - 2 \Omega g_{t\phi} -\Omega^2 g_{\phi\phi}}} \, ,
\ee
where in the expression of $\Omega$ the sign $+$ is for corotating orbits and the sign $-$ is for counterrotating orbits. We note that Eq.~(\ref{eq-NT}) gives exactly the same temperature profile as \citet{1974ApJ...191..499P} and can be applied to any stationary, axisymmetric, asymptotically flat, and circular spacetime.

Assuming that the disk is in thermal equilibrium, we can define an effective temperature $T_{\rm eff}$ at any radial coordinate of the disk by imposing $\mathcal{F} = \sigma T_{\rm eff}^4$, where $\sigma$ is the Stephan-Boltzmann constant. Since the temperature of the inner part of the accretion disk can be high in a black hole X-ray binary, non-thermal effects like electron scattering in the disk atmosphere are non-negligible. These non-thermal effects are taken into account by introducing the color factor $f_{\rm col}$ and defining the color temperature $T_{\rm col} = f_{\rm col} T_{\rm eff}$. The local specific intensity of the radiation emitted by the disk is
\be
I_{\rm e} \left( E_{\rm e} \right) = \frac{E_{\rm e}^3}{2 \pi^2} \frac{1}{f^4_{\rm col}}
\frac{\Upsilon}{\exp\left(\frac{E_{\rm e}}{T_{\rm col}}\right) - 1} \, ,
\ee
where $E_{\rm e}$ is the photon energy as measured in the rest-frame of the material in the disk and $\Upsilon$ is a function of the angle measured in the rest-frame of the material in the disk between the propagation direction of the photon emitted by the disk and the normal to the disk surface ($\Upsilon = 1$ for isotropic emission, which is the default value in our model and the case assumed in the next section). For a 10~$M_\odot$ black hole accreting at 10\% of the Eddington limit, $f_{\rm col}$ is expected to be in the range 1.5 to 1.9 (1.7 is our default value, which will be used even in the next section) \citep{1995ApJ...445..780S}. In the Kerr spacetime, the thermal emission at every radial coordinate $r$ is thus completely determined by $M$, $\dot{M}$, the black hole spin parameter $a_*$, $f_{\rm col}$, and $\Upsilon$.

The total luminosity of the source in Eddington units is $\lambda = L_{\rm tot}/L_{\rm Edd}$, where $L_{\rm tot}$ is the total luminosity of the source and $L_{\rm Edd}$ is its Eddington luminosity. Within our model, the total luminosity of the source is the sum of the contribution from the accretion disk and of that from the corona
\be
L_{\rm tot} = L_{\rm disk} + L_{\rm corona} = (1 + \rho) L_{\rm disk} \, ,
\ee
where we have defined $\rho = L_{\rm corona}/L_{\rm disk}$. The luminosity of the accretion disk $L_{\rm disk}$ can be easily calculated from the equation 
\be
L_{\rm disk} = \eta \dot{M} c^2 \, ,
\ee 
where $\eta = 1 - E_{\rm in}$ is the radiative efficiency of the system and $E_{\rm in}$ is the conserved specific energy in Eq.~(\ref{eq-energy}) calculated at the radius of the inner edge of the disk $r_{\rm in}$. In the next section, we present the predictions of {\tt ziji} in different regimes and we use the parameter $\lambda$ to regulate the total luminosity of the source and the parameter $\rho$ to regulate the relative strength between the luminosity of the disk and the luminosity of the corona. In such a context, the mass accretion rate $\dot{M}$ is not an independent parameter and is instead determined by the values of $M$, $\lambda$, and $\rho$.

\begin{table*}
\centering
\renewcommand\arraystretch{1.5}
\begin{tabular}{lcl}
\hline\hline
Parameter & Symbol & Notes \\
\hline
Black hole mass & $M$ & Black hole mass in units of $M_\odot$ \\
Black hole spin & $a_*$ & Dimensionless black hole spin parameter \\
Distance of the source & $D$ & Distance of the source in kpc (for the normalization of the  spectrum) \\
Viewing angle & $i$ & Inclination angle of the disk with respect to the line of sight of the observer in deg \\
Color factor & $f_{\rm col}$ & Default value 1.7 \\
Disk emission & $\Upsilon$ & Default function $\Upsilon = 1$ (isotropic emission) \\
Luminosity of the source & $\lambda$ & Total luminosity of the source in Eddington units: $\lambda = L_{\rm tot}/L_{\rm Edd}$ \\
Luminosity of the corona & $\rho$ & Relative luminosity of the corona: $\rho = L_{\rm corona}/L_{\rm disk}$ \\
Height of the corona & $h$ & Height of the lamppost corona in gravitational radii \\
Photon index & $\Gamma$ & Photon index of the Comptonized spectrum from the corona: $dN/dE \propto E^{-\Gamma}$ \\
High-energy cutoff & $E_{\rm cut}$ & High-energy cutoff of the Comptonized spectrum from the corona in keV \\
Inner edge of the disk & $r_{\rm in}$ & Radial coordinate of the inner edge of the disk in gravitational radii ($r_{\rm in} \ge r_{\rm ISCO}$) \\
Outer edge of the disk & $r_{\rm out}$ & Radial coordinate of the outer edge of the disk in gravitational radii \\
Disk electron density & $n_{\rm e}$ & Electron density of the disk in units of electrons~cm$^{-3}$. Default function $n_{\rm e} = {\rm constant}$ \\
Iron abundance & $A_{\rm Fe}$ & Iron abundance in units of the Solar iron abundance \\
\hline \hline   
\end{tabular}
\caption{\rm Summary of the parameters in the simplest version of {\tt ziji}. \label{t-para}}
\vspace{0.5cm}
\end{table*}

As default coronal geometry, we use the lamppost model, where the corona is a point-like source with isotropic emission along the black hole spin axis and is completely specified by the coronal height $h$. This is certainly the simplest choice. However, it is straightforward to consider more complicated coronal geometries with non-isotropic emission \citep{2022ApJ...925...51R}. A ray-tracing code calculates how the corona (direct radiation from the corona) and the disk (thermal radiation of the disk returning to the disk) illuminate the disk \citep{2024ApJ...965...66M}. The latter is characterized by an electron density profile $n_{\rm e} = n_{\rm e} (r)$ (which can be specified by the user, but in the next section we will consider the simplest case of constant electron density) and its iron abundance $A_{\rm Fe}$. The ionization parameter of the disk $\xi$ at the radial coordinate $r$ is determined by the total incident photon flux $\Phi_X$ and the electron density at that radial coordinate
\be\label{eq-xi}
\xi = \frac{4 \pi \Phi_X}{n_{\rm e}} \, .
\ee 
The details of these calculations have been already presented in \citet{2024ApJ...965...66M} and have been already discussed in the literature; see, for example, \citet{2019MNRAS.488..324I}, \citet{2022MNRAS.514.3965D}, and for a more pedagogical and detailed discussion \citet{2024arXiv240812262B}.

For the calculation of the reflection spectrum in the rest-frame of the material in the disk, we use the {\tt reflionx} code \citep{2005MNRAS.358..211R}. We do not use the {\tt reflionx} table because we want to calculate the reflection radiation with the correct spectrum of the incident radiation, while in the {\tt reflionx} table it is assumed that the radiation illuminating the disk has a power law spectrum (which is a good approximation only for the direct radiation from the corona). After calculating the reflection spectrum from the direct radiation from the corona and the returning radiation of the thermal spectrum, the model includes the effect of the returning radiation of the reflection spectrum itself with a few iterations\footnote{{\tt ziji} determines the final reflection spectrum in 4~iterations. As shown in \citet{2024ApJ...965...66M}, there are only negligible differences if we go beyond the fourth iteration.}. 
The details of these calculations have been already presented in \citet{2024ApJ...965...66M}. At the end, we have a thermal component from the Novikov-Thorne accretion disk, a Comptonized component from the corona, and a reflection component calculated with the actual spectrum of the radiation illuminating the disk (direct radiation from the corona and returning radiation from the disk). Table~\ref{t-para} lists the parameters of the model in its simplest version.


\section{Results}\label{s-res}

In this section, we present the predictions of {\tt ziji}. We set the black hole mass to $10~M_\odot$. To maximize the impact of the returning radiation, we assume that the black hole spin parameter is $a_* = 0.998$ and that the inner edge of the accretion disk is at the ISCO radius, $r_{\rm in} = r_{\rm ISCO}$. There is no emission from the plunging region $r < r_{\rm in}$. As shown in Fig.~1 in  \citet{2024ApJ...965...66M}, a significant fraction of the radiation emitted by the disk returns to the disk only in the very inner part of the accretion disk, when $r < 2~r_{\rm g}$, where $r_{\rm g} = M$ is the black hole gravitational radius. Already at $r = 2~r_{\rm g}$, most of the emitted radiation can escape to infinity and this fraction approaches 1 as the emission radius increases. We assume a lamppost corona and we set $h = 5~r_{\rm g}$, where $h$ is the coronal height (i.e., the radial coordinate in Boyer-Lindquist coordinates of the point-like corona along the black hole spin axis).

\begin{figure*}[t]
\begin{center}
\includegraphics[width=0.95\linewidth]{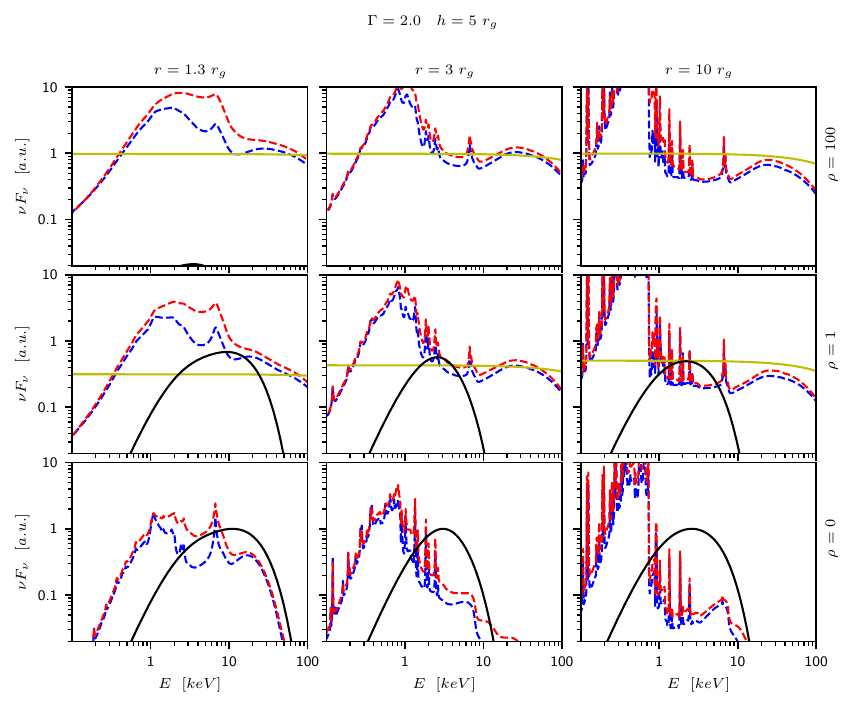}
\end{center}
\vspace{-0.5cm}
\caption{Synthetic reflection spectra in the rest-frame of the material in the disk at different radial coordinates ($r = 1.3$, 3, and 10~$r_{\rm g}$) when the Eddington-scaled luminosity of the source is $\lambda = 0.01$. Yellow solid curves describe the direct incident spectra from the corona, black solid curves describe the incident spectra of the returning radiation of the thermal component, blue dashed curves are the reflection spectra produced by the direct radiation from the corona and the returning radiation of the thermal component, and red dashed curves are for the final reflection spectra including the returning radiation of the reflection radiation. In the top panels, the spectrum of the source is dominated by the spectrum from the corona ($\rho = L_{\rm corona}/L_{\rm disk} = 100$). In the central panels, the luminosity of the corona and of the disk are equivalent ($\rho = 1$). In the bottom panels, there is no corona and the reflection spectrum is entirely generated by the returning radiation of the thermal component ($\rho = 0$). See the text for more details. \label{f-NR1}}
\end{figure*}

\begin{figure*}[t]
\begin{center}
\includegraphics[width=0.95\linewidth]{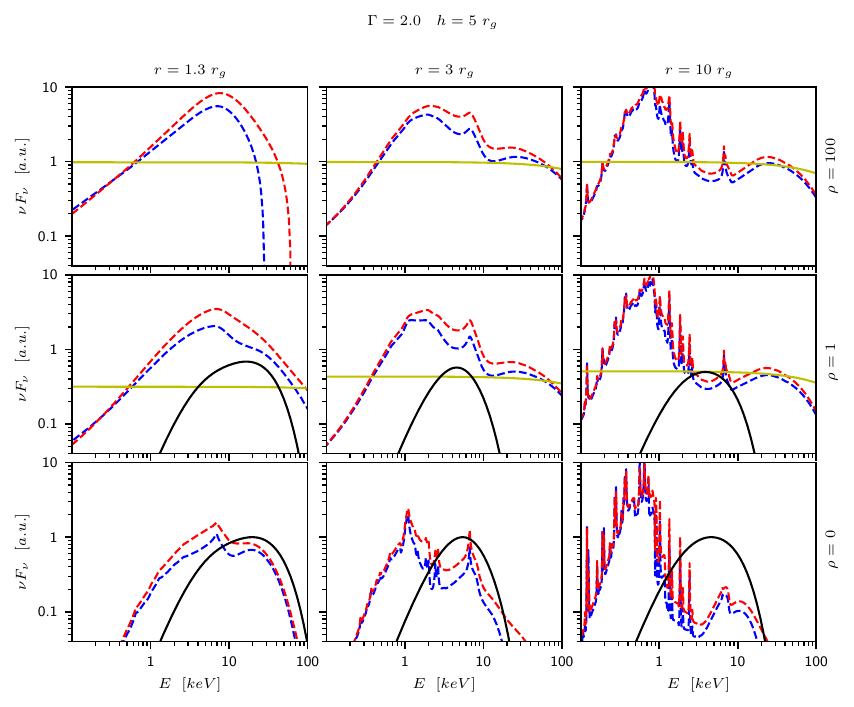}
\end{center}
\vspace{-0.5cm}
\caption{As in Fig.~\ref{f-NR1} for $\lambda = 0.1$. \label{f-NR2}}
\vspace{1.2cm}
\end{figure*}

The spectrum of the corona is a power law with an exponential high-energy cutoff $E_{\rm cut}$
\be
\frac{dN_{\rm c}}{dt_{\rm c} dE_{\rm c}} = K \, E_{\rm c}^{-\Gamma} \exp \left( - \frac{E_{\rm c}}{E_{\rm cut}} \right) \, ,
\ee
where the subindex ${\rm c}$ is used to indicate that a quantity is measured in the rest-frame of the corona, $K$ is a normalization constant, and we set the photon index $\Gamma$ to 2 and the high-energy cutoff $E_{\rm cut}$ to 300~keV. The total luminosity of the corona is
\be\label{eq-Lcorona}
L_{\rm corona} = \int_{E_{\rm min}}^{E_{\rm max}} E_{\rm c} \, \frac{dN_{\rm c}}{dt_{\rm c} dE_{\rm c}} \, dE_{\rm c} \, ,
\ee
where we set $E_{\rm min} = 1$~eV and $E_{\rm max}$ to infinity. Eq.~(\ref{eq-Lcorona}) is used to evaluate $\rho = L_{\rm corona}/L_{\rm disk}$, which can be regulated by the normalization constant $K$. The spectrum of the corona observed far from the source is redshifted. The redshift factor is  
\be
g = \left[ \sqrt{- g_{tt}} \right]_{r = h , \theta = 0} \approx 0.77 \, ,
\ee
where $g_{tt}$ is evaluated at the point of the corona, where $h = 5~r_{\rm g}$ in these simulations.

In Fig.~\ref{f-NR1}, we show the spectra in the rest-frame of the material in the disk at the radial coordinates $r = 1.3~r_{\rm g}$ (left panels), $3~r_{\rm g}$ (central panels), and $10~r_{\rm g}$ (right panels) when the Eddington-scaled luminosity of the source is $\lambda = 0.01$. The top panels are for $\rho = 100$ (the spectrum of the source is essentially the Comptonized spectrum from the corona), the central panels are for $\rho = 1$ (the corona and the disk have the same luminosity), and the bottom panels are for $\rho = 0$ (there is no corona and the reflection spectrum is generated exclusively by the returning radiation of the thermal spectrum of the disk). The electron density of the disk is $n_{\rm e} = 10^{22}$~cm$^{-3}$ at every radial coordinate and the ionization parameter $\xi$ is calculated from Eq.~(\ref{eq-xi}). The iron abundance is $A_{\rm Fe} = 1$. In every panel, we show the incident spectrum of the direct radiation from the corona (yellow solid line), the incident spectrum of the returning radiation of the thermal component (black solid line), the reflection spectrum generated by these two components (blue dashed line), and the final reflection spectrum that includes even the returning radiation of the reflection spectrum (red dashed line).

Fig.~\ref{f-NR2} is like Fig.~\ref{f-NR1} but for $\lambda = 0.1$. The higher luminosity of the source implies a higher ionization parameter and the spectra in the rest-frame of the material in the disk show fewer emission lines if compared to the panels in Fig.~\ref{f-NR1} with the same radial coordinate $r$. This is in particular the case for the panels for $r = 1.3$~$r_{\rm g}$, because the ionization parameter is too high to have emission lines.

\begin{figure*}[t]
\begin{center}
\includegraphics[width=0.95\linewidth]{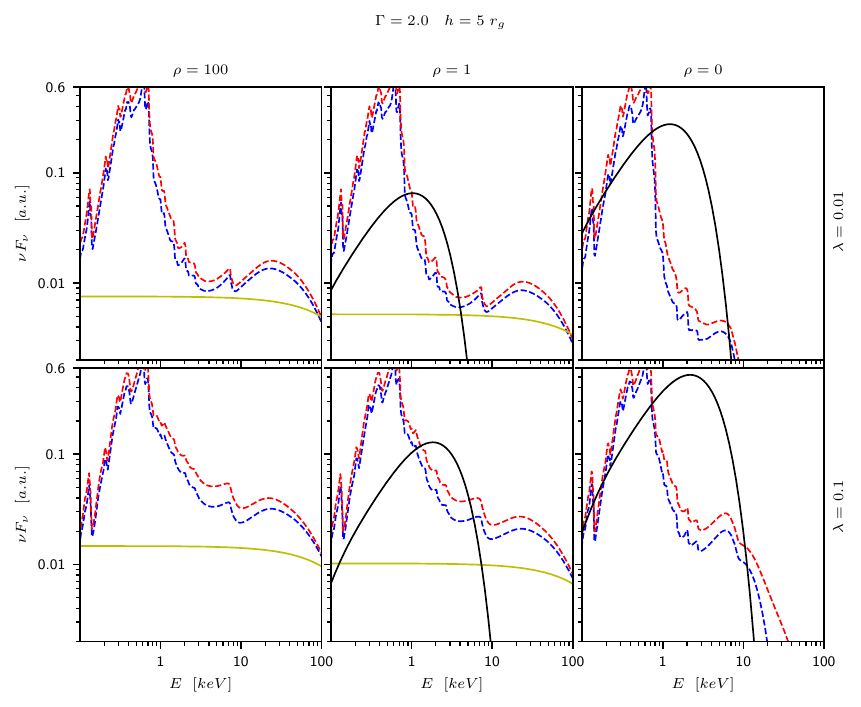}
\end{center}
\vspace{-0.5cm}
\caption{Synthetic reflection spectra of the whole disk as observed far from the source. The inclination angle of the disk with respect to the line of sight of the observer is $i = 45^\circ$. Yellow solid curves describe the Comptonized spectra from the corona. Black solid curves describe the thermal spectra of the disk. Blue dashed curves describe the reflection spectra produced by the direct radiation from the corona and the returning radiation of the thermal spectrum of the disk (i.e., we do not include higher order disk reflection). Red dashed curves are for our final results for the reflection spectra (i.e., we include higher order disk reflection). In the top panels, the Eddington-scaled luminosity of the source is $\lambda = 0.01$. In the bottom panels, we have $\lambda = 0.1$. In the left panels, the luminosity of the source is dominated by the corona ($\rho = L_{\rm corona}/L_{\rm disk} = 100$). In the right panels, there is no corona and therefore the reflection spectrum is produced by the returning radiation of the thermal component ($\rho = 0$). In the central panels, the luminosity of the corona and of the disk are equivalent ($\rho = 1$). See the text for more details. \label{f-R}}
\end{figure*}

\begin{figure*}[t]
\begin{center}
\includegraphics[width=0.95\linewidth]{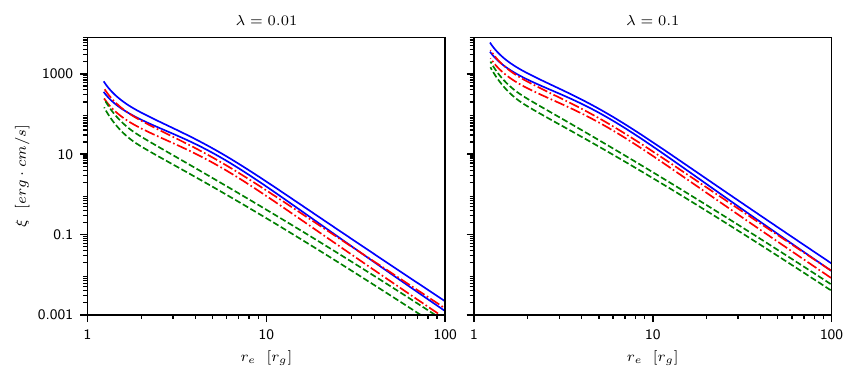}
\end{center}
\vspace{-0.5cm}
\caption{Radial profile of the ionization parameter $\xi$ in the simulations shown in Fig.~\ref{f-R}. The Eddington-scaled luminosity of the source is $\lambda = 0.01$ (left panel) and $\lambda = 0.1$ (right panel). Dashed-green curves are for $\rho = L_{\rm corona}/L_{\rm disk} = 0$, dashed-dotted-red curves are for $\rho = 1$, and solid-blue curves are for $\rho = 100$. The curves in the same style in the same panel refer to the ionization parameter calculated, respectively, without including (lower curves) and including (upper curves) the returning radiation of the reflection spectrum. \label{f-ion}}
\vspace{0.8cm}
\end{figure*}

\begin{figure*}[t]
\begin{center}
\includegraphics[width=0.95\linewidth]{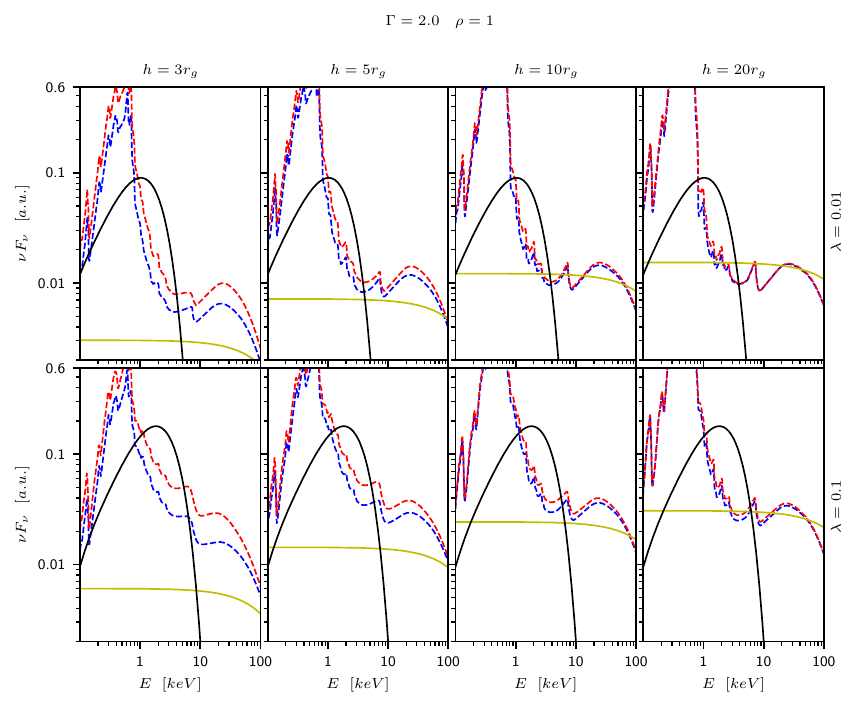}
\end{center}
\vspace{-0.5cm}
\caption{As in Fig.~\ref{f-R} for $\rho = 1$ and different values of the coronal height $h$. From left to right: $h = 3$~$r_{\rm g}$, 5~$r_{\rm g}$, 10~$r_{\rm g}$, and 20~$r_{\rm g}$. \label{f-height}}
\end{figure*}

Fig.~\ref{f-R} shows the reflection spectra of the whole disk as observed far from the source. The radial profile of the ionization parameter $\xi$ in these simulations is shown in Fig.~\ref{f-ion}. If we considered a lower disk electron density $n_{\rm e}$, like the default value $10^{15}$~cm$^{-3}$ in {\tt xillver}, which is the most widely used value in current reflection studies, the inner part of the accretion disk would be fully ionized. We also note that $\rho$ is not the only parameter determining the relative contribution between the corona and the returning radiation of the thermal component in the generation of the reflection spectrum. The height of the corona $h$ also plays an important role. The impact of the coronal height on the reflection spectrum observed far from the source is shown in Fig.~\ref{f-height}, where we assume $\rho = 1$ and we only change the value of $h$.

\section{Discussion and conclusions}\label{s-dis}

In the previous section we have presented the predictions of {\tt ziji} assuming the source is a 10~$M_\odot$ black hole for two different Eddington-scaled luminosities, $\lambda = 0.01$ and 0.1. For both luminosities, we have considered three different regimes: the spectrum of the source is dominated by the spectrum of the corona (roughly corresponding to a hard state), the spectrum of the source is dominated by the thermal spectrum of the disk (roughly corresponding to a soft state), and the spectrum of the source is the fair combination of the Comptonized spectrum of the corona and the thermal spectrum of the disk (roughly corresponding to an intermediate state).

We note that the spectra presented in the previous section have been obtained assuming the disk electron density $n_{\rm e} = 10^{22}$~cm$^{-3}$, which is the maximum value allowed by the {\tt reflionx} code. For lower values of the electron density, the relativistic reflection spectra do not present strong reflection features, as the ionization of the accretion disk is very high. This would not be consistent with observations, where we can see strong reflection features even in the X-ray spectra of Galactic black holes with Eddington-scaled luminosities of 0.3 and more. As shown in Fig.~\ref{f-ion}, $\xi > 10$~erg~cm~s$^{-1}$ in the inner part of the accretion disk even for $\lambda = 0.01$. If we employed the disk electron density $n_{\rm e} = 10^{15}$~cm$^{-3}$, which is still the most widely used value in current reflection studies, the disk would be fully ionized and the reflection spectrum would have no clear emission lines. This point is often overlooked in the literature because in current reflection models there is no link between the reflection spectrum and the Eddington luminosity of the source, but actually current reflection models would be suitable to fit only the data of black hole binaries with extremely low Eddington luminosity. High density reflection models seem thus strictly necessary if we want to fit the spectra of black hole X-ray binaries \citep{2023ApJ...951..145L}.

We also note that our model does not include the returning radiation in the calculation of the thermal spectrum of the disk. As shown in \citet{2005ApJS..157..335L}, the returning radiation has the effect to heat the accretion disk and the thermal spectrum calculated including the returning radiation is very similar to a thermal spectrum calculated without the returning radiation but with a higher value of the mass accretion rate. In the analysis of the spectra of a source we can measure an effective mass accretion rate, which is slightly higher than the actual mass accretion rate because the calculations do not include the effects of the returning radiation.

The relativistic reflection spectra calculated for $\rho = 0$ (no corona) do not present any Compton hump, see the right panels in Fig.~\ref{f-R}. This could have been expected, as the thermal spectrum of the disk is peaked in the soft X-ray band. It can be an observational signature for reflection spectra entirely or predominantly generated by the returning radiation of the thermal component. We note that a weak Compton hump has been already observed in the spectra of the black hole X-ray binary 4U~1630 and interpreted as a reflection spectrum generated by the returning radiation of the thermal component of the disk in \citet{2021ApJ...909..146C}.

If we compare the reflection spectra for $\rho = 100$ and for $\rho = 1$ in Fig.~\ref{f-R}, we do not see clear differences. This means that the the contribution of the returning radiation of the thermal spectrum in the generation of the reflection spectrum is still subdominant when $\rho = 1$. The reflection spectra with $\rho = 0$ are instead clearly different from those with $\rho = 1$. The role of the returning radiation of the thermal spectrum is certainly important when a source is in a soft state, but it can be presumably ignored during intermediate and hard states. A quantitative analysis on when the contribution of the returning radiation of the thermal spectrum is significant is beyond the scope of this work and does not depend only on the value of $\rho$. As shown in Fig.~\ref{f-height}, even the coronal height $h$ can affect the strength of the Compton hump. The characteristics of the X-ray detectors and of the source are also quite important. The Compton hump region seems to be the most sensitive part of the spectrum and therefore an observation in which we can measure the Compton hump well is probably suitable to measure the parameter $\rho$ and distinguish a source in which the reflection spectrum is mainly produced by the illumination of the disk by a corona from a source in which the reflection spectrum is mainly generated by the returning radiation of the thermal component.

The astrophysical model behind {\tt ziji} is quite sophisticated and its predictions should be more accurate than current reflection models used for data analysis~\citep{2004ApJS..153..205D,2014ApJ...782...76G,2019MNRAS.485.2942N,2019MNRAS.488..324I,2019ApJ...878...91A}. However, the model is definitively too slow to be directly used to analyze real data and there are too many parameters (even in the minimal version presented in the previous section) to construct a model with FITS files. If we consider current reflection models used for data analysis, they calculate a reflection spectrum by solving an integral. Inside the integral, there is the reflection spectrum in the rest-frame of the material in the disk (non-relativistic reflection spectrum) and the transfer function, which takes into account all relativistic effects and transforms the reflection spectrum in the rest-frame of the material in the disk into the reflection spectrum of the whole disk as detected far from the source (relativistic reflection spectrum); see, e.g., \citet{2017ApJ...842...76B}. Both the non-relativistic reflection spectra and the transfer functions are pre-calculated and tabulated into FITS files, so during the data analysis process the model calls these FITS files and quickly calculate the integral. In this way it is possible to generate many relativistic reflection spectra for different values of the model parameters and find the best-fit model. The {\tt reflionx} table has around five parameters (depending on the exact version): the photon index $\Gamma$ and the high-energy cutoff $E_{\rm cut}$ describing the spectrum of the corona, and the electron density $n_{\rm e}$, the ionization parameter $\xi$, and the iron abundance $A_{\rm Fe}$ to describe the accretion disk. The returning radiation adds some broad features to the direct radiation from the corona and such broad features can be approximated by introducing 5-10 parameters into the model. A {\tt reflionx} table with 10-15 parameters would require the construction of a large FITS file, which would make the model too slow. However, we could replace the FITS file of the {\tt reflionx} table with a neural network. The model could call the neural network to predict the non-relativistic reflection spectra and then solve the integral. For example, we could use a variational autoencoder like that described in \citet{2023arXiv231204640S} for the calculations of black hole images. In our case, the variational autoencoder should predict the non-relativistic reflection spectra for a model with 10-15~parameters.



{\bf Acknowledgments --} 
This work was supported by the National Natural Science Foundation of China (NSFC), Grant No.~12250610185 and 12261131497, and the Natural Science Foundation of Shanghai, Grant No.~22ZR1403400.
T.M. also acknowledges the support from the China Scholarship Council (CSC), Grant No.~2022GXZ005433.
J.J. acknowledges support from Leverhulme Trust, Isaac Newton Trust and St Edmund's College, University of Cambridge.
S.R. also acknowledges the support from the Teach@T\"ubingen Fellowship.



\end{document}